\documentclass{emulateapj}
\usepackage{epstopdf}
\usepackage{graphicx}
\usepackage{longtable}

\newcommand {\aplt} {\ {\raise-.5ex\hbox{$\buildrel<\over\sim$}}\ }

\slugcomment{Accepted for Publication in ApJ Letters}
\shorttitle{ASASSN-15\lowercase{oi}}
\shortauthors{Gezari et al.}

\begin{document}

\title{X-ray Brightening and UV Fading of Tidal Disruption Event ASASSN-15\lowercase{oi}}

\author{S. Gezari\altaffilmark{1,2}, S. B. Cenko\altaffilmark{3,2}, and I. Arcavi\altaffilmark{4,5,6}}
\altaffiltext{1}{Department of Astronomy, University of Maryland, Stadium Drive, College Park, MD 20742-2421, USA \email{suvi@astro.umd.edu}}
\altaffiltext{2}{Joint Space-Science Institute, University of Maryland, College Park, MD 20742, USA}
\altaffiltext{3}{Astrophysics Science Division, NASA Goddard Space Flight Center, Mail Code 661, Greenbelt, MD 20771, USA}
\altaffiltext{4}{Department of Physics, University of California, Santa Barbara, CA 93106-9530, USA}
\altaffiltext{5}{Las Cumbres Observatory Global Telescope, 6740 Cortona Dr., Suite 102, Goleta, CA 93117-5575, USA}
\altaffiltext{6}{Einstein Fellow}

\begin{abstract}
We present late-time observations by {\it Swift} and {\it XMM-Newton} of the tidal disruption event (TDE) ASASSN-15oi  that reveal that the source brightened in the X-rays by a factor of $\sim10$ one year after its discovery, while it faded in the UV/optical by a factor of $\sim 100$.   The {\it XMM-Newton} observations measure a soft X-ray blackbody component with $kT_{\rm bb} \sim 45$ eV, corresponding to radiation from several gravitational radii of a central $\sim 10^6 M_\odot$ black hole.  The last {\it Swift} epoch taken almost 600 days after discovery shows that the X-ray source has faded back to its levels during the UV/optical peak.   The timescale of the X-ray brightening suggests that the X-ray emission could be coming from delayed accretion through a newly forming debris disk, and that the prompt UV/optical emission is from the prior circularization of the disk through stream-stream collisions.   The lack of spectral evolution during the X-ray brightening disfavors ionization breakout of a TDE ``veiled'' by obscuring material.   This is the first time a TDE has been shown to have a delayed peak in soft X-rays relative to the UV/optical peak, which may be the first clear signature of the real-time assembly of a nascent accretion disk, and provides strong evidence for the origin of the UV/optical emission from circularization, as opposed to reprocessed emission of accretion radiation.  \end{abstract}


\section{Introduction}

The tidal disruption of a star by a central supermassive black hole (SMBH) is expected to result in a flare of radiation from the accretion of the bound debris via a newly formed accretion disk \citep{Rees1988, Ulmer1999}.  The characteristic temperature of a circularized disk of stellar debris accreting onto a supermassive black hole (SMBH) with $R_{\rm disk}=2R_{\rm T}$ (as expected from angular momentum conservation) is $T_{\rm max} = 4.2\times10^{5}~{\rm K}~M_{6}^{-1/4}$ or $kT_{\rm max} = 36~{\rm eV}~M_{6}^{-1/4}$ (Miller 2015), where $M_{6}=M_{\rm BH}/(10^{6} M_{\odot}$) and $R_{\rm T} \sim R_{\star}(M_{\rm BH}/M_{\star})^{1/3}$ is the tidal disruption radius.  This temperature corresponds to a spectral peak in the soft X-rays at  $\sim 0.2$ keV.  

Indeed, the first candidates for TDEs were discovered in the soft X-ray (0.1-2.4 keV) band by the \textsl{ROSAT} All-Sky Survey.  \textsl{ROSAT} detected luminous X-ray outbursts from several apparently inactive galaxies whose extremely soft spectra, dramatic fading on the timescale of years, and rate of $\approx 10^{-4}$ per year per galaxy were consistent with the theoretical expectations for TDEs \citep{Komossa2002, Donley2002}.  Several more soft X-ray candidates were subsequently discovered with the \textsl{XMM-Newton} Slew Survey and in searches of archival \textsl{Chandra} data \citep{Esquej2008, Maksym2010, Saxton2017}, with a range of blackbody temperatures ($kT_{\rm bb} = 0.04-0.12$ keV).

However, dedicated searches with wide-field UV and optical surveys have found a population of candidate TDEs with surprisingly low blackbody temperatures on the order of $\sim 1-3 \times 10^{4}$ K \citep{Gezari2008, vanVelzen2011, Gezari2012, Arcavi2014, Holoien2014, Holoien2016, Holoien2016-14li, Blagorodnova2017, Hung2017}, at odds with the basic predictions of radiation from a newly formed accretion disk in a TDE.  These lower temperatures have been attributed to larger radii associated with a reprocessing layer \citep{Loeb1997, Guillochon2014, Roth2016}, potentially formed from a radiatively driven wind \citep{Metzger2015, Strubbe2015, Miller2015}, or radiation from stream-stream collisions in the circularizing debris disk itself  \citep{Lodato2012, Piran2015, Shiokawa2015, Jiang2016, Krolik2016, Bonnerot2017, Wevers2017}.  

Another surprise has been the X-ray weakness of these TDEs discovered in the UV and optical surveys.  Only a few optical TDE candidates have been detected with X-ray emission in follow-up observations: \textsl{GALEX} D1-9 and \textsl{GALEX} D3-13 \citep{Gezari2008}; ASASSN-14li \citep{Holoien2016-14li}; ASASSN-15oi \citep{Holoien2016}, with upper limits in a few cases: PS1-10jh \citep{Gezari2012}; iPTF16fnl \citep{Blagorodnova2017}; iPTF16axa \citep{Hung2017, Brown2018}.  

ASASSN-15oi is a TDE reported by \citet{Holoien2016} that was discovered by the ASAS-SN survey on 2015 August 14 in the nucleus of a quiescent early-type galaxy at $z=0.0484$ ($d_L = 216$ Mpc for $H_0 = 69.6$ km s$^{-1}$ Mpc$^{-1}$, $\Omega_M = 0.29, \Omega_\Lambda = 0.71$).  It was notable for its significant temperature evolution in the first two months after discovery, increasing from $2 \times 10^{4}$ K to $4 \times 10^{4}$ K, and a relatively faint soft X-ray flux.  Otherwise, its peak luminosity ($L \sim 1.3 \times 10^{44}$ erg s$^{-1}$), power-law decay of its light curve, and He-dominated optical spectrum, were similar to other optically discovered TDEs.  Here we report the late-time X-ray brightening of the TDE ASASSN-15oi, and show that these observations are consistent with the delayed onset of accretion.  This indicates that the optical/UV emission in this TDE (and perhaps by association others) is due to circularization of the debris, rather than re-processed accretion radiation, as suggested by \citet{Piran2015}. The paper is organized as follows.  We present the new observations by {\it Swift} and {\it XMM-Newton} in \S 2, the evolution of the X-ray spectrum, and UV and X-ray light curves up to 600 days after discovery in \S \ref{sec:lc}, a comparison of the relevant timescales of a TDE to the timescale of the dramatic evolution in the UV-to-X-ray ratio observed in ASASSN-15oi, and our conclusions for the nature of the X-ray and UV/optical components in \S \ref{sec:disc} and \S \ref{sec:conc}.

\section{Observations} \label{sec:obs}
\subsection{Swift Observations} \label{sec:swift}

ASASSN-15oi was monitored with {\it Swift} for $\sim 40$ epochs in all six UVOT filters: UVW2 (1928 \AA), UVM2 (2246 \AA), UVW1 (2600 \AA), U (3465 \AA), B (4392 \AA), and V (5468 \AA) between 2015 Aug 27 and  2016 July 24 with a typical exposure time of $\sim 2.5$ ks.  We requested one more epoch with UVW2 filter on 2017 April 1.  We performed aperture photometry using a 5$\arcsec$ radius aperture and a 20$\arcsec$ background region using the {\tt uvotsource} task in HEASoft \footnote{https://heasarc.gsfc.nasa.gov/lheasoft/}.  The measured magnitudes are reported in the AB system, and have not been corrected for the host galaxy flux or Galactic extinction.  We analyzed the simultaneous observations with {\it Swift} XRT \citep{Burrows2005} in the 0.3-10 keV band with standard XRT analysis procedures \citep[e.g.,][]{Evans2009}, and convert from count-rate to absorbed flux using a factor of $3.0 \times 10^{-11}$ erg s$^{-1}$ cm$^{-1}$ (cps)$^{-1}$ appropriate for a $kT_{\rm bb} \sim 0.04$ keV blackbody (see Table \ref{tab2}).  

\begin{figure*}
\includegraphics[angle=0,scale=.5,trim={0 0 0 0},clip]{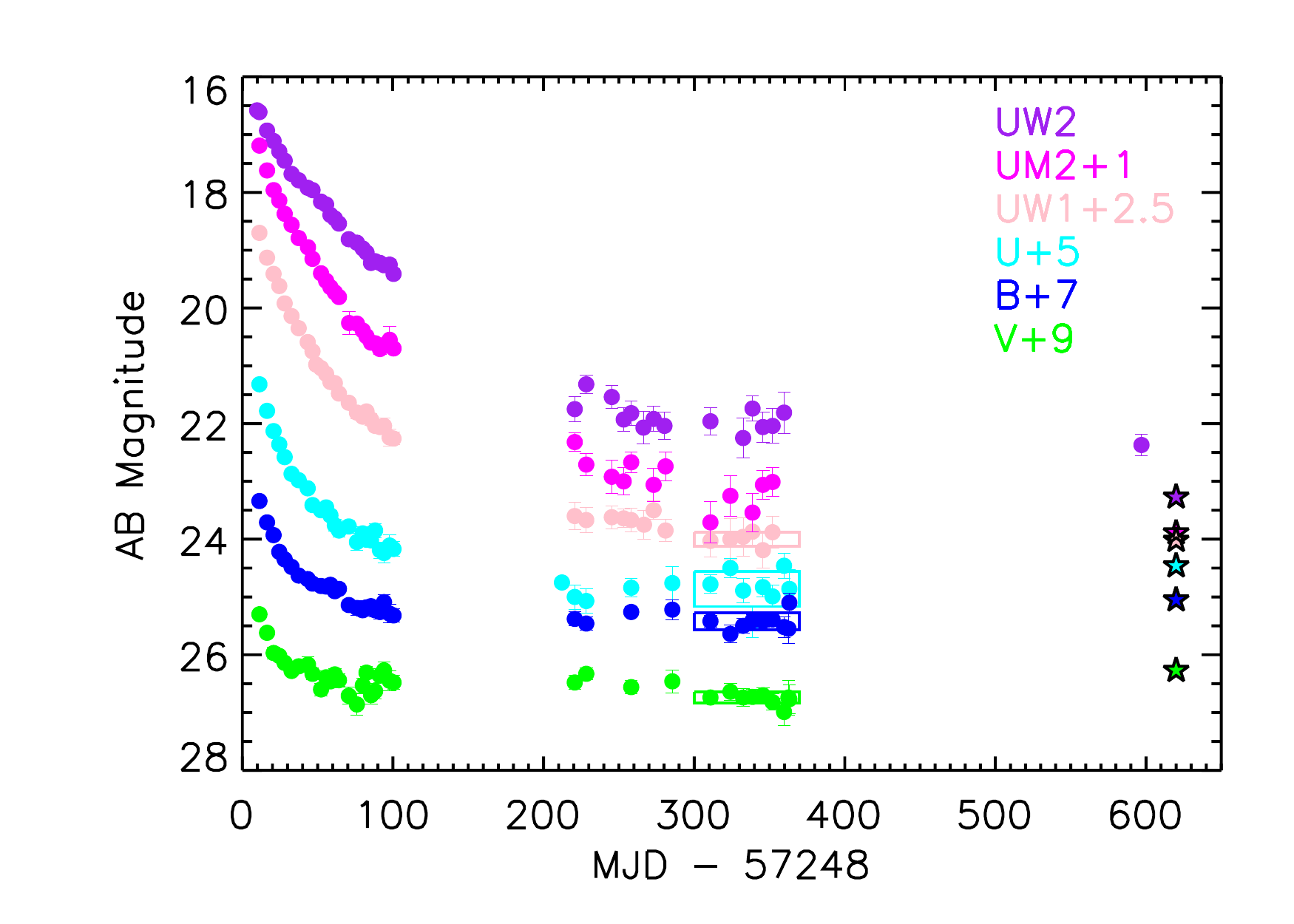}
\includegraphics[angle=0,scale=.5, trim={0 0 0 0},clip]{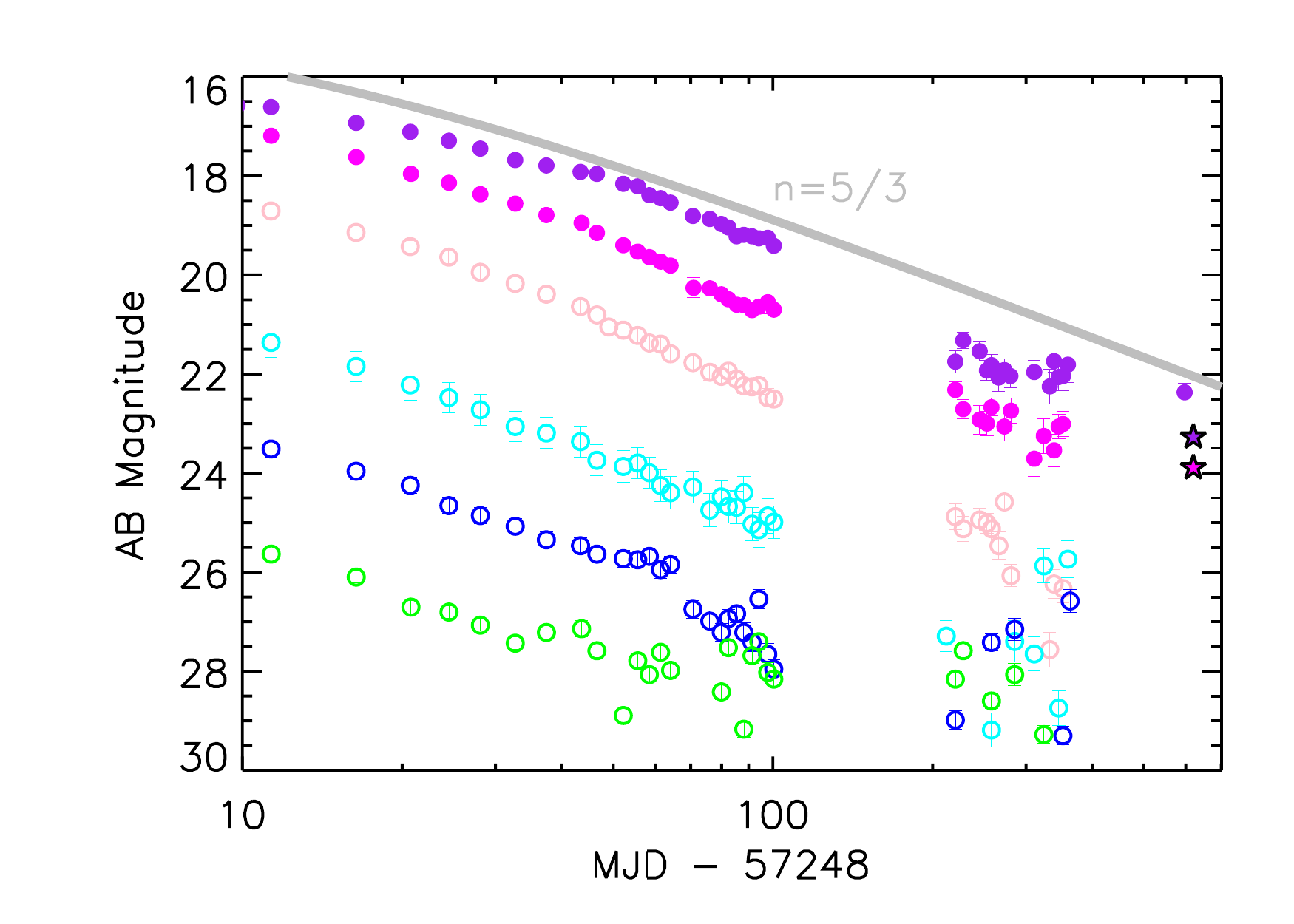}
\caption{
Light curve of ASASSN-15oi measured by {\it Swift} UVOT.  On the {\it left} we indicate the plateau in flux measured at $\Delta t > 300$ days after discovery in the UVW1, U, B, and V filters, suggesting that the flux is dominated by the host galaxy flux in these filters at late times.  On the {\it right} we plot the UVOT light curve after subtracting off the plateau flux level in the UVW1, U, B, and V filters to account for the host galaxy contribution, in comparison to a $t^{-n}$ power-law decay, were $n=5/3$.  In both figures, stars represent the synthetic magnitudes for the host galaxy presented in \citet{Holoien2016} from a stellar population fit to the \textsl{GALEX} $NUV$ upper limit, ASAS-SN $V$-band, and 2MASS $JHK_{S}$ archival magnitudes of the host galaxy.  \label{fig:uvot} }
\end{figure*}

In the optical {\it Swift} UVOT filters, the light curves plateau, indicating that the transient has likely faded below the host galaxy flux.  We estimate the host galaxy flux from the plateau at $\Delta t > 300$ days after discovery to be UVW1=$21.50 \pm 0.1$, U=$19.86 \pm 0.3$, B=$18.42 \pm 0.2$, and V=$17.74 \pm 0.1$ mag, respectively.  
The last UVW2 detection at $t=597$ days is UVW2$=22.37 \pm 0.18$ mag, which is below the archival \textsl{GALEX} upper limit of $NUV > 22.0$ mag, but almost 1 mag brighter than the synthetic UVOT host galaxy magnitude reported in \citet{Holoien2016} of UVW2$=23.27 \pm 0.13$ mag.  Thus we assume that the UVW2 flux is still dominated by the transient.  In Figure \ref{fig:uvot} we show the observed {\it Swift} UV/optical light curve, as well as the light curve with the host flux (estimated from the observed plateau flux level) subtracted off in the UVW1, U, B, and V filters.  After this host flux subtraction, all six filters have a similar power-law decline with little evidence of color evolution.

\subsection{XMM-Newton Observations }\label{sec:obs}
We triggered two epochs of TOO {\it XMM-Newton} 17 ksec observations (PI: Gezari) on 2015 October 29 and 2016 April 4 with the EPIC-pn 0.2-12 keV detector in Full Frame mode with the thin filter.   The data were reduced with the XMM-Newton Science Analysis System (XMM-SAS) software package version 15.0.  After filtering for background flaring, we were left with 12.1 ks (MOS1), 12.5 ks (MOS2), 10.3 ks (pn) of usable exposure time in 2015 and 15.5 ks (MOS1), 15.4 ks (MOS2), and 12.0 ks (pn) in 2016.  We measured the X-ray flux in a 300 pixel aperture, with a background region measured in an annulus with a radius of 500 and 1500 pixels, respectively.  We detect a total of 75 (MOS1), 69 (MOS2), 453 (pn) counts in 2015 and 417 (MOS1), 386 (MOS2), and 2882 (pn) counts in 2016.    Finally, we bin in energy to yield a minimum of 3 counts per bin.  Given the much larger number of counts in the pn channel, we present our spectral analysis using the pn spectra.

\section{Analysis} \label{sec:lc}
\subsection{X-ray Brightening}

The 2015 {\it Swift} data reported in \citet{Holoien2016} from 2015 Aug 29 - Nov 17 was fitted with two components, a soft blackbody with $kT = 49 \pm 9$ eV  and a $\Gamma=1.76 \pm 1.0$ power-law and with Galactic absorption fixed to $N_H = 5.59\times 10^{20}$ cm$^{-2}$, and a total flux in the 0.3-10 keV band not corrected for Galactic absorption of $f_{\rm abs}$ (0.3-10 keV) = $(8.0 \pm 0.2) \times 10^{-14}$ erg s$^{-1}$ cm$^{-2}$.  The X-ray flux was constant within the errors during this monitoring period, and the {\it ROSAT} archival upper limit ($f (0.3-10)$ keV $< 1.2 \times 10^{-13}$ erg s$^{-1}$ cm$^{-2}$) was not constraining enough for \citet{Holoien2016} to determine if this X-ray emission was indeed associated with the UV/optical transient.  

\begin{figure*}
\includegraphics[angle=-90,scale=.3,trim={0 0 40 0},clip]{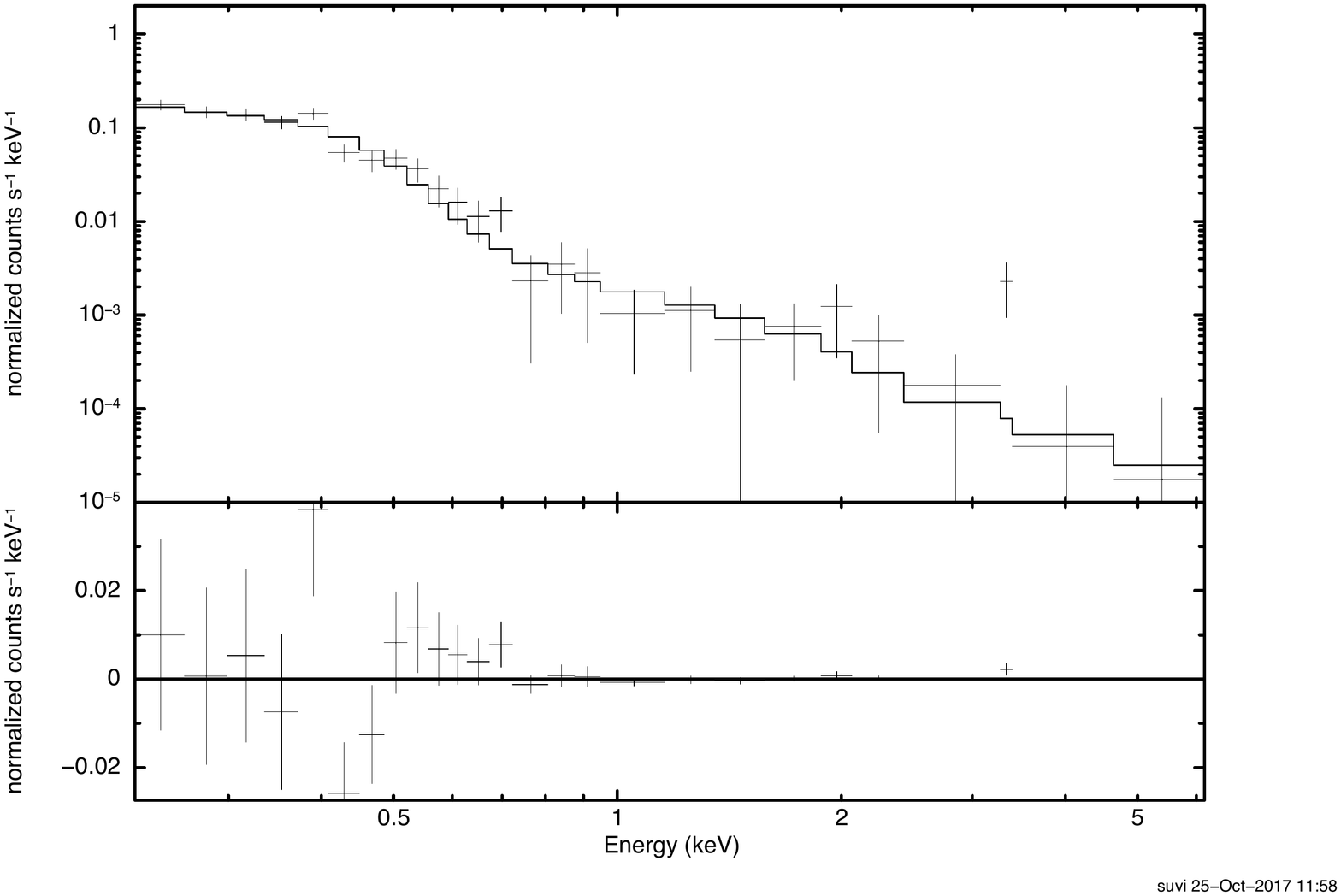}
\includegraphics[angle=-90,scale=.3, trim={0 0 40 0},clip]{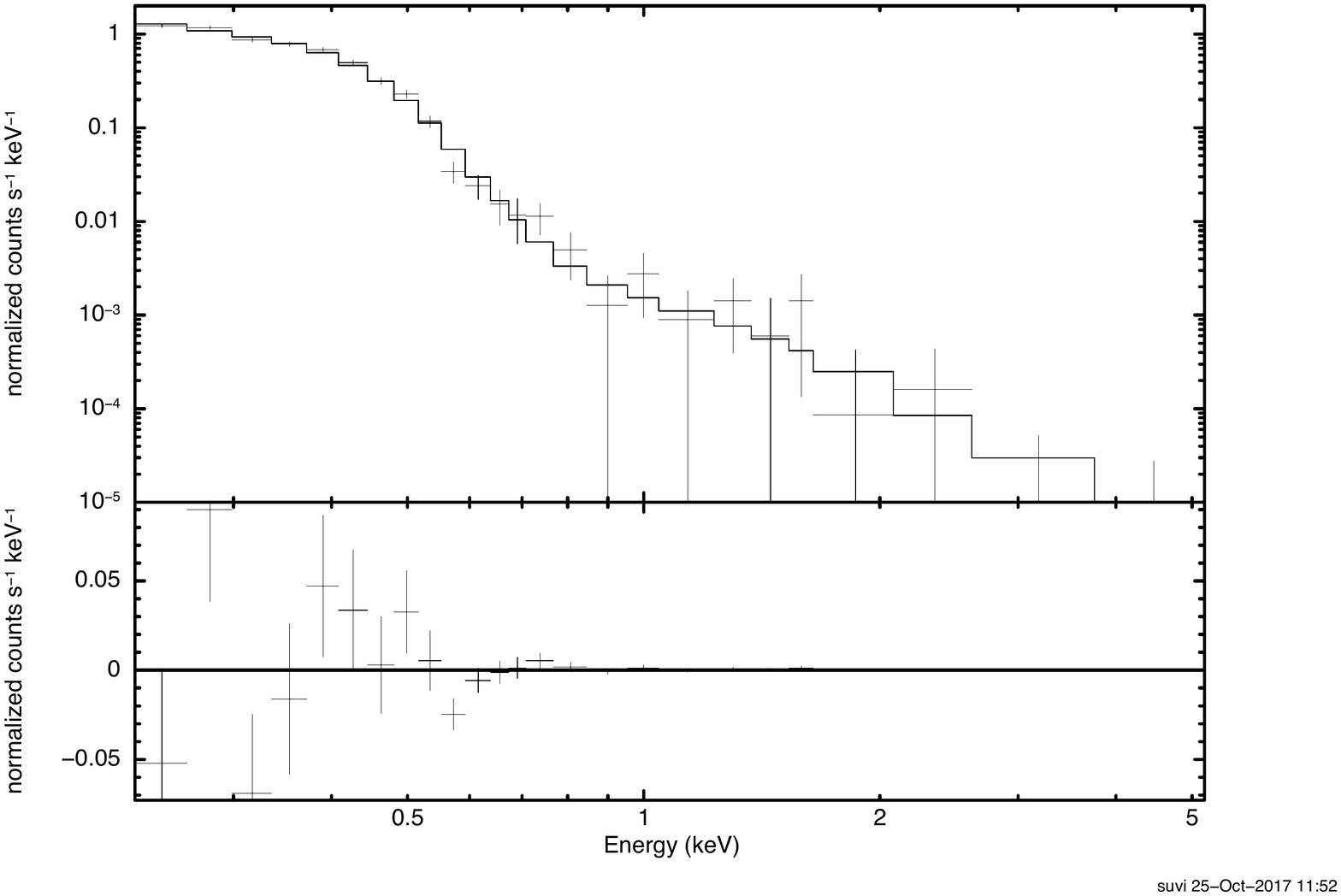}
\caption{
{\it XMM-Newton} pn spectra for ASSASN-15oi including a model fit and residuals for a blackbody plus power law and Galactic absorption from 2015 October 29 ({\it Left}) and 2016 April 4 ({\it Right}), demonstrating a factor of $12 \pm 1$ increase in the soft blackbody component.  The parameters, normalizations (in units of of photons s$^{-1}$ cm$^{-2}$ keV$^{-1}$), and reduced chi-square per degree of freedom ($\chi_{\rm dof}$) of the fit in 2015 are: $kT = 47 \pm 3$ eV, $A_{\rm bb}=(2.1 \pm 0.5) \times 10^{-5}$  and $\Gamma = 2.5 \pm 0.8$, $A_{\rm pl} =(2.6 \pm 0.7) \times 10^{-6}$, with $\chi_{\rm dof}^2 = 1.02$; and in 2016 are: $kT = 42 \pm 0.7$ eV, $A_{\rm bb} = (2.5 \pm 0.2) \times 10^{-4}$ and $\Gamma = 3 \pm 1$, $A_{\rm pl} = (2.0 \pm 0.7) \times 10^{-6}$, with $\chi_{\rm dof}^2 = 1.21$. \label{fig:xmm} }
\end{figure*}

Our first {\it XMM-Newton} epoch on 2015 October 29 was obtained during the {\it Swift} monitoring observations reported in \citet{Holoien2016}.   Using the X-ray spectral fitting package XSPEC version 12.9.0, the spectrum is very well fitted with a blackbody plus power-law ($\chi_{\rm dof}^2 = 1.02$), with $kT_{\rm bb} = 47.4 \pm 2.5$ eV and $\Gamma = 2.5 \pm 0.8$ with $f_{\rm abs}$ (0.3-10 keV) = $6.5 \times 10^{-14}$ erg s$^{-1}$ cm$^{-2}$.  The blackbody temperature and power-law index are both consistent within the errors of the model fit measured from stacking the {\it Swift} XRT observations in 2015.  The second {\it XMM-Newton} epoch was taken on 2016 April 4, 234 days after the optical transient was discovered by the ASAS-SN survey.  The spectrum is again well described with the blackbody plus power-law model ($\chi_{\rm dof}^2 = 1.21$), with $kT_{\rm bb} = 42.3 \pm 0.7$ eV and $\Gamma = 3.3 \pm 1.3$ and with a total absorbed flux of $f_{\rm abs}$ (0.3-10 keV) = $3.7 \times 10^{-13}$ erg s$^{-1}$ cm$^{-2}$.  Note that while the power-law component normalization remains constant within the errors between the two epochs, the blackbody component normalization increases by a factor of $12 \pm 1$ (see Figure \ref{fig:xmm}).  

In the last {\it Swift} XRT observation on 2017 April 4 we detect a source with a count rate consistent with the {\it Swift} count-rate measured in 2015.  The brightening of the X-ray emission in 2016 thus places a lower limit on the X-ray flux associated with ASASSN-15oi of $f_{\rm abs}$ (0.3-10 keV) $> 3 \times 10^{-13}$ erg s$^{-1}$ cm$^{-2}$.  The constant  power-law ($\Gamma \sim 2.5$) component of the X-ray emission may be associated with an underlying low-luminosity AGN with $L(0.3-10$ keV) = $6.4 \times 10^{40}$ erg s$^{-1}$, for $d_{L} = 215$ Mpc and correcting for Galactic absorption.  However, the extremely variable and soft blackbody component is most likely originating from the TDE, with a luminosity for the blackbody component corrected for Galactic absorption of$L(0.3-10)$ keV = $9 \times 10^{41}$ erg s$^{-1}$ at $\Delta t = 76$ days since discovery increasing to $L(0.3-10)$ keV = $1 \times 10^{43}$ erg s$^{-1}$ at $\Delta t = 234$ days since discovery.  The corresponding bolometric luminosity for the soft blackbody component is $9.4 \times 10^{42}$ erg s$^{-1}$ and $1.9 \times 10^{44}$ erg s$^{-1}$, corresponding a radius of $3.8 \times 10^{11}$ cm and $2.1 \times 10^{12}$ cm, respectively, a factor of 5.5 increase in radius.  The late-time decline in X-rays is consistent with what is expected for emission on the Wien's tail of the thermal TDE emission, $L_{\rm X} \propto$exp($At^{-5/12}$), where $A$ is a constant that depends on the parameters of the event \citep{Lodato2011}, suggesting that after 1 year, the X-rays are now following more closely the fallback rate.

\begin{figure*}
\plotone{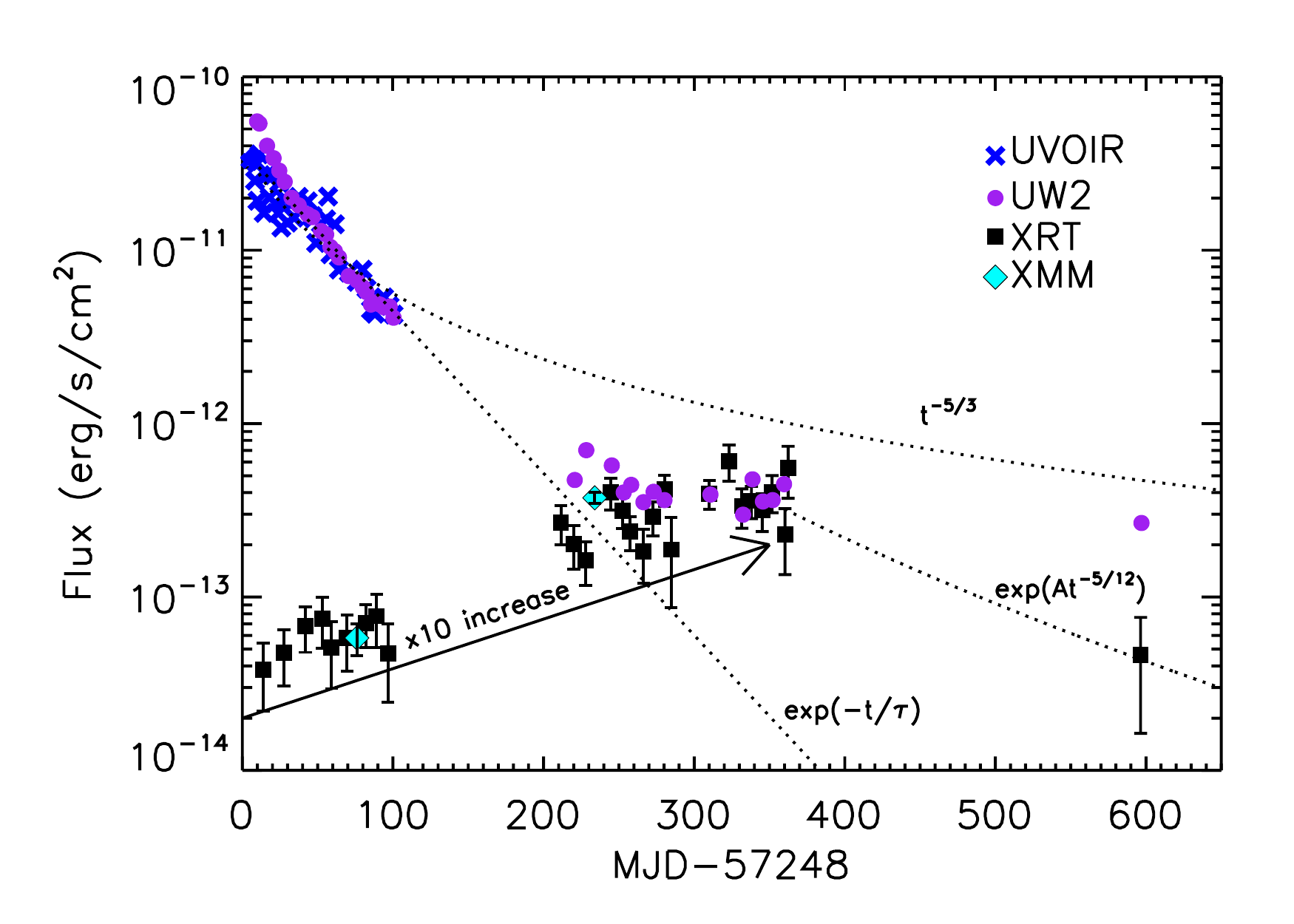}

\caption{
X-ray light curve of ASASSN-15oi measured by {\it XMM-Newton} (XMM: cyan diamonds with 1 $\sigma$ error bars) and {\it Swift} XRT (black squares with 1 $\sigma$ error bars), in comparison to the blackbody fit to the UV/optical component reported in \citet{Holoien2016} (UVOIR: blue X's), and in our extended UV monitoring in the UVW2 band with {\it Swift} UVOT (UW2: purple dots).  The UV/optical light declines more slowly than an exponential decline, and more steeply than a $t^{-5/3}$ decline.  The X-ray light curve brightens by a factor of 10 over 1 year, and then declines back to pre-event levels, at a rate consistent with that expected for X-ray emission on the Wien's tail of the thermal disk emission.  \label{fig:xrt} }
\end{figure*}

The {\it Swift} XRT and {\it XMM-Newton} observations follow the brightening of the absorbed X-ray flux by a factor of 10 from $\sim 5 \times 10^{-14}$ erg s$^{-1}$ cm$^{-2}$ to $\sim 5 \times 10^{-13}$ erg s$^{-1}$ cm$^{-2}$ over a time period of 1 year (see Figure \ref{fig:xrt}).  This is in stark contrast to the fading of the UV flux by a factor of 100 over this same time period.  This difference in temporal behavior, combined with the much larger inferred radius of the UV/optical emission from blackbody fits of $10^{14}-10^{15}$ cm \citep{Holoien2016} suggests that these components are physically distinct.

\begin{figure*}
\plotone{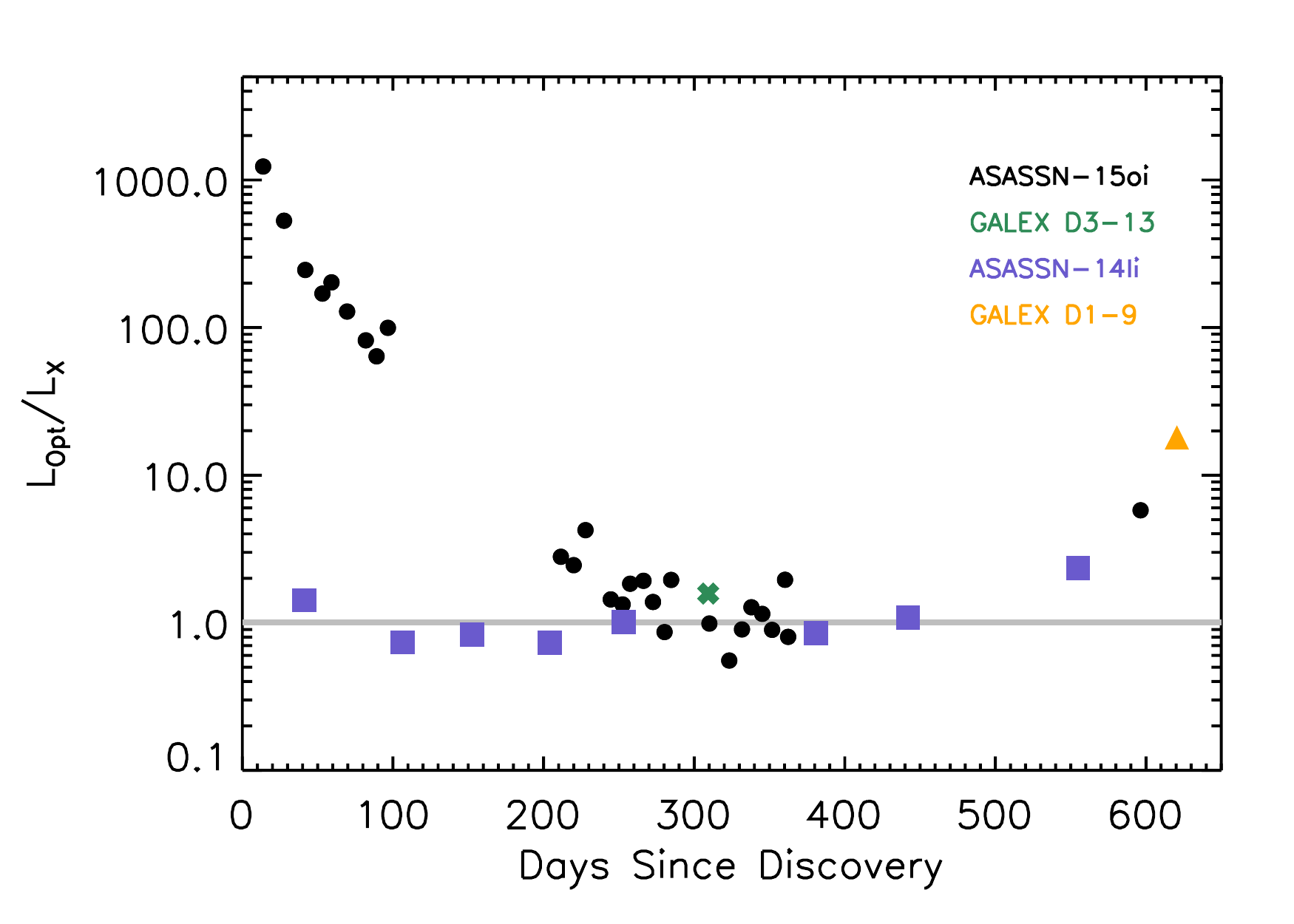}
\caption{
Compilation of UV/optical to X-ray flux ratio for TDEs with both components detected.   Grey line shows the value of $L_{\rm opt}/L_{X} \sim 1$ that appears to be characteristic of well-studied TDE ASASSN-14li.  \label{fig:ratio} }
\end{figure*}

\subsection{UV Fading}

  In Figure \ref{fig:xrt} we plot the total UV/optical flux by scaling the UVW2 flux density to match the bolometric flux measured from a blackbody fit to the UV/optical blackbody component 50-100 days after discovery reported by \citet{Holoien2016}.  The UV/optical light curve decline at late times is shallower than the exponential decline with $\tau = 46.5$ d fitted by \citet{Holoien2016} from the first 100 days of {\it Swift} monitoring, and is steeper than the $t^{-5/3}$ power-law decay consistent expected for the bolometric luminosity evolution in a TDE \citep{Phinney1989, Lodato2009, Guillochon2013}.  When we fit the light curve with a $t^{-n}$ power-law, we find $n=3 \pm 1$, but any single power-law model is unable to fit the plateau in UV/optical flux at late-times.

\subsection{Evolution of the Optical to X-ray Ratio}

In Figure \ref{fig:ratio} we plot $L_{\rm opt}/L_{\rm X}$ in days since discovery for ASASSN-14li and ASASSN-15oi \citep{Holoien2016-14li, Holoien2016}, as well as for two \textsl{GALEX} TDEs that had X-ray detections in their late-time follow-up {\it Chandra} observations: D3-13 and D1-19 \citep{Gezari2006, Gezari2008}.  We do not plot TDEs with X-ray upper limits, such as PS1-10jh \citep{Gezari2012}, ASASSN-14ae \citep{Holoien2014}, iPTF16axa \citep{Hung2017}, and iPTF-16fnl \citep{Blagorodnova2017}, since the X-ray band is on the Wien's tail of the thermal emission expected from TDEs, and it is not clear if a non-detection in the 0.3-10 keV band of the X-ray detectors is due to a low flux or a very soft spectrum.  For example, if the temperature of ASASSN-15oi had been just a little cooler, $kT_{\rm bb} < 30$ eV, the flux density at 0.3 keV would have been a factor of 50 fainter.

In the first year of monitoring of ASASSN-15oi there is a dramatic decrease in the UV/optical to X-ray (0.3-10 keV) luminosity ratio ($L_{\rm opt}/L_{\rm X}$) from $\sim 1000$ to $\sim 1$.  This is in stark contrast to the relatively constant $L_{\rm opt}/L_{\rm X} \sim 1$ observed in ASASSN-14li, the only other non-jetted TDE with a well sampled X-ray light curve.  And yet, the blackbody temperature and inferred radius for ASASSN-14li are very similar to that measured for the soft X-ray component of ASASSN-15oi, with $kT_{\rm bb} = $51 eV and $r_{\rm BB}= 1.7 \times 10^{12}$ cm =  $11 M_6^ {-1} r_{\rm g}$ \citep{Miller2015}.    

\section{Discussion} \label{sec:disc}

In order to compare the evolution of ASASSN-15oi to the relevant timescales for a TDE, we must first constrain its central black hole mass.  \citet{Holoien2016} estimate $M_{\rm BH} \sim 10^{7.1} M_\odot$ from its host galaxy mass of $10^{10.0 \pm 0.1} M_\odot$ (consistent within the errors from the estimate of \citet{vanVelzen2017} of $10^{9.9} M_\odot$), and assuming a bulge-to-total mass ratio of 0.575 and the $M_{\rm BH}-M_{\rm bulge} $ relation from \citet{McConnell2013}.  

However, we note that in the study of the velocity dispersion ($\sigma_\star$) of TDE host galaxies by \citet{Wevers2017}, they find that the inferred $M_{\rm BH}$ in TDEs are systematically higher when using the $M_{\rm BH}-M_{\rm bulge}$ relation than when using the $M_{\rm BH}-\sigma_\star$ relation.  Similarly, AGN are found to have black hole masses that lie an order of magnitude below the $M_{\rm BH}-M_{\rm bulge}$ relation established from dynamical studies of local galaxies with larger black hole masses than the AGN samples ($> 10^8 M_\odot$) \citep{Reines2015}.  \citet{Reines2015} do find a scaling relation between total stellar mass and black hole mass for AGN with $10^{5} M_\odot < M_{\rm BH} < 10^{8} M_\odot$ that would imply a black hole mass for the host galaxy of ASASSN-15oi of only $10^{(6.4 \pm 0.55)} M_\odot$.  Given the downward trend in the inferred black hole mass for ASASSN-15oi, we scale our equations to a $10^{6} M_\odot$ black hole.  

The characteristic timescale for a TDE is set by the orbital period of the most tightly bound debris, known as the fallback time ($t_{\rm fb}$), which for a solar-type star is:

\[ t_{\rm fb} = 41~{\rm d}~M_6^{1/2}. \]

\noindent The circularization timescale ($t_{\rm circ}$) driven by relativistic apisidal precession of the debris streams depends on the black hole mass as

\[t_{\rm circ} = 8.3 t_{\rm fb}~M_6^{-5/3} \beta^{-3} \]

where $\beta = R_{\rm T}/R_{\rm p}$ \citet{Bonnerot2016}.   Meanwhile, the viscous inflow time scale for a standard $\alpha$-disk model \citep{Shakura1973} is

\[t_{\rm visc} = \alpha^{-1} (h/r)^{-2} P_{\rm out} \sim 0.1 t_{\rm fb} (\alpha/0.1)^{-1} (h/r)^{-2}\]

\noindent where $\alpha$ is the standard viscous parameter, $h$ is the scale-height of the disk, and $P_{\rm out} $ is the orbital period of the outer edge of the disk.  

In the case of inefficient circularization, there will be a ``viscous delay" between the fallback time and the onset accretion of the debris though a disk.   In \citet{Guillochon2015} they predict a population of ``viscously delayed" TDEs from lower-mass back holes ($M_{\rm BH} \aplt 10^6 M_{\odot}$) with weaker stream-stream collisions as a consequence of weaker relativistic precession, that result in longer timescales and lower accretion rates.  The circularization timescale ($t_{\rm circ}$) for a $10^{6} M_\odot$ black hole is tantalizingly close to the $\sim 1$ year rise-time to peak observed in the soft X-rays for ASASSN-15oi, and may indicate that we are indeed seeing delayed accretion due to inefficient circularization of the debris disk. The blackbody radii inferred from the {\it XMM-Newton} observations are indeed consistent with the inner parts of an accretion disk, with $r_{\rm bb} = 3 M_6^ {-1} r_{g}$ on 2015 October 29 and $r_{\rm bb} = 15 M_6^ {-1} r_{\rm g}$ on 2016 April 4, where $r_{g} = GM/c^2 = 1.5 \times 10^{11} M_6$ cm, and the slow expansion rate ($\sim$ 1 km s$^{-1}$) may be tracing the viscous spreading of the newly formed accretion disk.   (Note that if the black hole mass in ASASSN-15oi is in fact closer to $10^7 M_\odot$, then the inferred circularization timescale is a factor of 15 times shorter, and its blackbody radius a factor of 10 smaller in units of $r_g$.)

Another scenario is that the accretion disk is enshrouded in material that is optically thick to soft X-rays.   In this case, the soft X-ray radiation can eventually ``break out" once the obscuring material expands enough to become fully ionized \citep{Metzger2016}.    The nature of this expansion could be attributed to a radiatively driven outflow, or from the evolution of the circularizing debris streams themselves, which are expected to form a plume of debris that expands due to energy dissipated from the shocked streams \citep{Jiang2016}.  
\citet{Auchettl2017} suggested that the ratio of X-ray to optical power observed in TDEs is related to the Eddington ratio, where TDEs with higher accretion rates have more material available to obscure the X-ray emission.  
In the case of ASASSN-15oi, since the soft blackbody component of the X-ray emission is detected at early times, it can only have been partially veiled.  Furthermore, the lack of evidence for a decrease in absorbing column density in ASASSN-15oi with the increasing X-ray luminosity argues against the ``veiled'' TDE scenario.

Finally, we can rule out that the X-ray properties of ASASSN-15oi are due to a viewing angle effect, since the orientation of the black hole and debris stream are not expected to change on the timescales observed for the X-ray to optical flux evolution in ASASSN-15oi.  

\section{Conclusions} \label{sec:conc}
We present the brightening by a factor of $\sim 10$ of the soft X-ray luminosity in TDE ASASSN-15oi measured by {\it Swift} and {\it XMM-Newton} on a timescale of 1 year after discovery.  The decoupled behavior of the brightening soft X-ray emission relative to the fading UV/optical emission suggest that they arise from physically distinct components.  The timescale and spectrum of the delayed soft X-ray peak is consistent with the circularization timescale and inner radius of a TDE debris disk around a $\sim 10^6 M_\odot$ black hole.   The prompt onset of the UV/optical emission is then naturally explained if it originates in shocks in the debris streams in the process of circularization \citep{Lodato2012, Piran2015}, and may apply to the nature of optical emission in TDEs in general.

\acknowledgements

We thank the referee for their helpful comments.  S.G. was supported in part by XMM-Newton grant NNX14AF36G and NSF CAREER grant 1454816.   The authors thank the {\it Swift} TOO team for promptly approving and executing our late-time observations.  Support for IA was provided by NASA through the Einstein Fellowship Program, grant PF6-170148.
\bibliographystyle{fapj}

\clearpage

\begin{deluxetable}{lcc}
\tablecaption{{\it Swift} XRT Photometry \label{tab2}}
\tablewidth{0pt}
\tablehead{
\colhead{MJD} & \colhead{Flux} & \colhead{Error} \\
\colhead{} & \colhead{($\times 10^{-14}$ erg s$^{-1}$ cm$^{-2}$)} & \colhead{($\times 10^{-14}$ erg s$^{-1}$ cm$^{-2}$)}
}
\startdata
57261.807 &  3.70 &  1.56 \\
57275.677 &  4.63 &  1.65 \\
57289.790 &  6.58 &  1.92 \\
57301.169 &  7.28 &  2.36 \\
57307.205 &  4.94 &  2.06 \\
57317.555 &  5.62 &  1.99 \\
57329.969 &  6.88 &  1.92 \\
57337.125 &  7.53 &  2.57 \\
57344.641 &  4.60 &  2.19 \\
57459.502 & 24.69 &  6.26 \\
57468.005 & 18.56 &  5.22 \\
57475.854 & 14.97 &  4.19 \\
57492.635 & 37.11 &  7.75 \\
57500.477 & 28.88 &  5.88 \\
57505.498 & 21.96 &  4.90 \\
57514.334 & 16.90 &  5.83 \\
57520.652 & 26.77 &  5.93 \\
57528.327 & 38.61 &  7.87 \\
57532.715 & 17.32 &  9.28 \\
57558.094 & 36.35 &  6.87 \\
57571.295 & 56.12 & 13.29 \\
57579.733 & 30.87 &  7.90 \\
57586.012 & 33.09 &  6.99 \\
57593.215 & 29.19 &  7.21 \\
57599.735 & 37.27 &  9.00 \\
57608.337 & 21.10 &  8.69 \\
57610.389 & 51.24 & 17.02 \\
57844.412 &  4.27 &  2.76 
\enddata
\end{deluxetable}

\end{document}